\documentstyle[psfig,twocolumn,prl,aps,amssymb]{revtex}
\begin{document}
\wideabs{
\draft

\title{Magneto-roton excitation of fractional quantum Hall effect: Comparison
between theory and experiment}
\author{Vito W. Scarola, Kwon Park, and Jainendra K. Jain}
\address{Department of Physics, 104 Davey Laboratory,
The Pennsylvania State University, University Park, Pennsylvania 16802}

\date{\today}

\maketitle

\begin{abstract}

A major obstacle 
toward a {\em quantitative} verification, by comparison to experiment,
of the theory of the excitations of the fractional quantum Hall effect 
has been the lack of a proper understanding of disorder.  We circumvent 
this problem by studying the {\em neutral} magneto-roton excitations, whose 
energy is expected to be largely insensitive to disorder.  The calculated 
energies of the roton at 1/3, 2/5 and 3/7 fillings of the lowest Landau level  
are in good agreement with those measured experimentally.

\end{abstract}

}


Quantitative tests of the theory of the fractional quantum Hall effect (FQHE)
have focused in the past primarily on the gap to 
charged excitation, determined experimentally 
from the temperature dependence of the longitudinal resistance.
A factor of two discrepancy between theory and experiment 
has persisted over the years, 
believed to be caused by disorder for which a quantitatively reliable 
theoretical treatment is not available at the moment.
In recent years there has been tremendous experimental progress in the 
measurement of the energy of the neutral magneto-roton excitation \cite {GMP}, 
both by inelastic  Raman scattering \cite{Pinczuk1,Pinczuk2,Davies} and 
by ballistic phonon absorption \cite {Mellor,Zeitler,Devitt}, and its 
energy has been determined at Landau level fillings of 1/3, 2/5, and 3/7.
While the neutral magneto-roton is of great interest in its own right,  
being the lowest energy excitation of the FQHE state, the 
chief motivation of this work is the observation that the disorder
is not likely to affect its energy significantly, in contrast to the energy 
of the charged excitation, 
because the roton has a much weaker dipolar coupling to disorder due to its
overall charge neutrality, and the coupling is further diminished because 
the disorder in modulation doped samples 
is typically smooth on the scale of the size (on the order a magnetic
length) of the spatially localized roton.
There is also compelling experimental evidence for the insensitivity of 
the roton energy to disorder: the same roton energy was found for samples 
for which the gaps in transport experiments 
differed by as much as a factor of two \cite {Mellor}.
The roton therefore provides a wonderful opportunity 
for testing the quantitative validity of
our understanding of the excitations of the fractional quantum Hall state.  
With this goal in mind, we have undertaken a comprehensive and 
realistic calculation of the roton energy at several filling factors in 
the lowest Landau level (LL).

The neutral excitation of the FQHE will be treated in the framework of the 
composite fermion (CF) theory \cite {Jain,Review1,Review2}, the composite 
fermion being the bound state of an electron and an even number of flux 
quanta (a flux quantum is defined as $\phi_0=hc/e$).  According to this theory, 
the interacting electrons at Landau level filling factor 
$\nu=n/(2pn\pm 1)$, $n$ and $p$ being integers, transform into weakly
interacting composite fermions at an effective filling $\nu^*=n$; 
the ground state corresponds to $n$ filled CF-LLs and 
the neutral excitation to a particle-hole pair of 
composite fermions, called the CF exciton.
Microscopic wave functions for the CF ground state and the CF exciton 
are readily constructed by analogy to the known wave functions of
the electron ground state at filling factor $n$, $\Phi^{gs}_{n}$, and its 
exciton, $\Phi_{n}^{ex}$:
\begin{equation}
\Phi_{\frac{n}{2n+1}}^{gs}={\cal P}_{LLL} \prod_{j<k}(z_j-z_k)^{2p}\Phi_n^{gs}
\end{equation}
\begin{equation}
\Phi_{\frac{n}{2n+1}}^{ex}={\cal P}_{LLL}\prod_{j<k}(z_j-z_k)^{2p}\Phi_n^{ex},
\end{equation}
where $z_j=x_j+i y_j$ is the position of the $j$th particle, 
and ${\cal P}_{LLL}$ 
denotes projection of the wave function into the lowest Landau level.
It was shown earlier that $\Phi_{\nu}$ can be obtained from $\Phi_n$ by
substituting the single electron wave functions $Y_{\alpha}({\bf r}_j)$ by
the `single CF wave functions' $Y^{CF}_{\alpha}({\bf r}_j)={\cal
P}_{LLL}\prod_{k}^{'}(z_j-z_k)^{p}Y_{\alpha}({\bf r}_j)$,
the explicit form of which has been given in the literature \cite{JK}.
(The prime denotes the condition $k\neq j$.) The composite fermion 
interpretation of $\Phi_{\nu}$ follows since multiplication by
the Jastrow factor $\prod_{j<k}(z_j-z_k)^{2p}$ is tantamount to attaching $2p$ 
flux quanta to each electron, converting it into a composite fermion.
These wave functions have been found to be quite accurate in tests against exact
diagonalization results available for small systems \cite {Jain,JK}.

The Hamiltonian for the many electron system is given by 
\begin{equation}
H=\frac{1}{2}\sum_{j\neq k}V(r_{jk})+V_{e-b}
\end{equation}
where $V_{e-b}$ is the electron-background interaction, with the background
assumed to be comprised of a uniform positive charge, and $V(r)$ is the
effective two-dimensional electron-electron interaction.  
(The kinetic energy is quenched in the lowest Landau level.)
For a strictly two-dimensional system, 
$V(r_{jk})=\frac{e^2}{\epsilon|r_j-r_k|},$ where $\epsilon$ is the dielectric
constant of the background material.  
As we will see, an important quantitative correction comes from the
finite transverse extent of the electron wave function, which 
alters the form of the effective two-dimensional interaction at short distances.  
The effective interaction can be calculated straightforwardly once the 
transverse wave function is known, which in turn will be determined by
self-consistently solving the Schr\"odinger and Poisson equations, taking
into account the interaction effects through the local density 
approximation (LDA) including the exchange correlation potential
\cite {Stern}.  Two geometries, single heterojunction 
and square quantum well (SQW), are considered due to their experimental 
relevance.  To simplify the calculation, we assume that
the electron wave function is confined entirely on the GaAs side of the
heterojunction, which is a reasonably good approximation for deep 
confinement.
It is stressed that neither the microscopic wave function nor the effective
interaction contains any adjustable parameters; the former depends  
only on the filling factor, while the latter is determined from a 
first principles, self-consistent LDA calculation, 
with the two-dimensional 
density, the sample type (heterojunction or square quantum well) and the 
known sample parameters  as the only input.

The energy of the exciton at $\nu=\frac{n}{2n+1}$,  
\begin{equation} 
\Delta^{ex} =\frac{<\Phi_{\nu}^{ex}|H|\Phi_{\nu}^{ex}>} 
{<\Phi_{\nu}^{ex}|\Phi_{\nu}^{ex}>} 
-\frac{<\Phi_{\nu}^{gs}|H|\Phi_{\nu}^{gs}>}{<\Phi_{\nu}^{gs}|\Phi_{\nu}^{gs}>}
\end{equation}  
is computed by Monte Carlo methods in the standard spherical geometry \cite
{Haldane}.
Since moving a single particle at each step of the Monte Carlo 
changes the single CF wave function $Y^{CF}$  
for {\em all} particles, the full 
wave function must be computed at each step.
The exciton wave function is a linear superposition of $\sim N/n$ Slater
determinants, but each of these  
differs from the ground state only in one row, and 
the clever techniques for upgrading Slater determinants \cite {Ceperley} 
significantly reduce the computing time, enabling us to study reasonably large
systems (up to 63 composite fermions were used in the present study). 
The ground and excited state
energies are evaluated sufficiently accurately to get a reasonable
estimate for the gap.
Due to the lack of edges in the spherical geometry being studied, we
expect that the gap will have a linear dependence  on $N^{-1}$ to leading
order, which is also borne out by our results.
A linear extrapolation to the thermodynamic limit $N^{-1}\rightarrow 0$ 
is taken after correcting the energies for the finite size deviation of the 
density from its thermodynamic value in the standard manner \cite{Morf}.
All results below are thermodynamic extrapolations, unless mentioned
otherwise.  The energies are quoted in units of
$e^2/\epsilon l_0$, where $l_0=\sqrt{\hbar e/eB}$ is the magnetic length.

\begin{figure}
\centerline{\psfig{figure=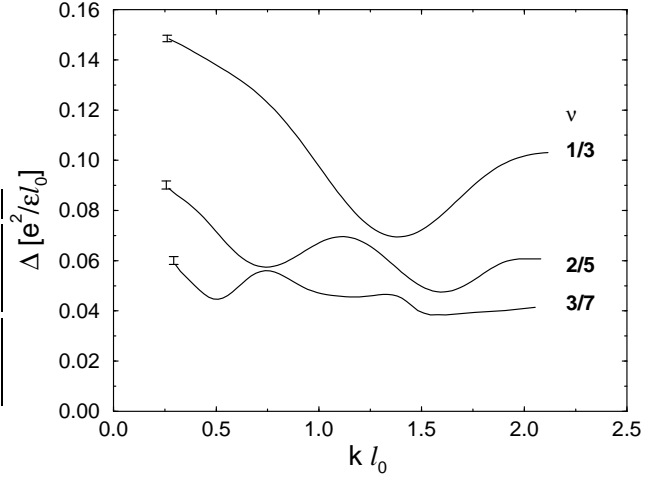,width=4.0in,angle=-90}}
\caption{The dispersions of the CF excitons at $\nu=1/3$, 2/5, and 
$\nu=3/7$ for a strictly two dimensional system.
The curves are obtained from the discrete points of 
systems with up to 50 particles (without extrapolation to 
the thermodynamic limit), with
the typical Monte Carlo uncertainty shown at the beginning of each curve.
The energies are given in units of $e^2/\epsilon l_0$, 
where $\epsilon =12.8$ is the
dielectric constant of GaAs, and $l_0$ is the magnetic length.
\label{fig1}}
\end{figure}

We have determined the dispersion of the CF exciton for 
1/3, 2/5, and 3/7, corresponding to one,
two, and three filled CF-LLs, respectively.
The typical dispersion contains several minima, as shown in Fig.~(\ref{fig1}).
We will term the lowest energy minimum the ``fundamental" roton, or simply the
roton, the others being secondary rotons.  Since only discrete values of $k$
are available in our finite systems, the energy of the roton is obtained by 
fitting the points near the minimum to a parabolic dispersion 
\begin{equation}
\Delta_k^{ex}=\Delta+\frac{\hbar^2 (k-k_0)^2}{2 m_R^*}
\label{mass}
\end{equation}
for each $N$, and then extrapolating $\Delta$ to the thermodynamic limit.
The energies in the $kl_0\rightarrow 0$ limit and at the roton minimum are
given in Table I for a strictly two dimensional system, along with $m_R^*$.

\begin{figure}
\centerline{\psfig{figure=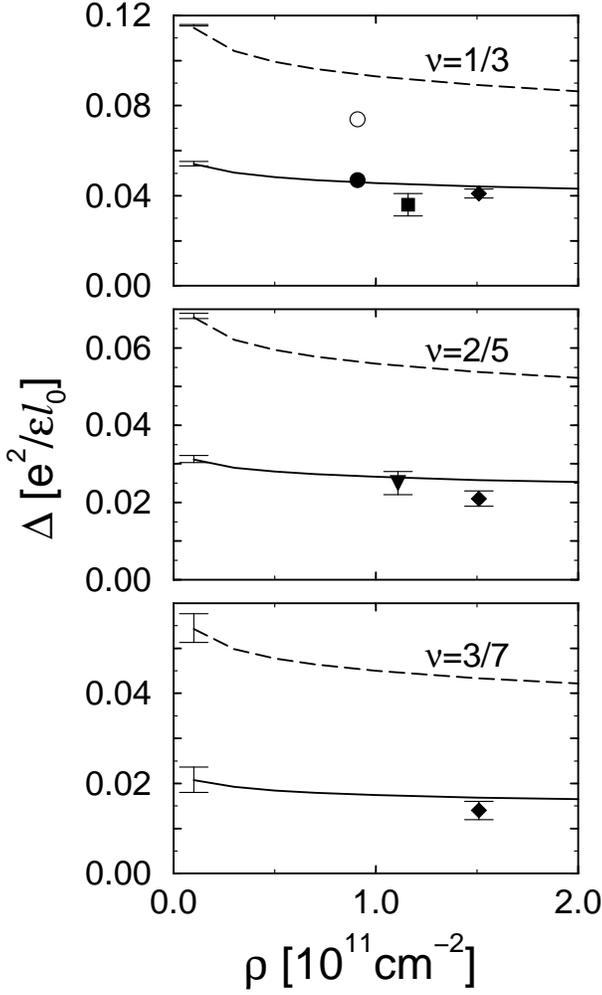,width=5.0in,angle=0}}
\caption{The energies of the CF roton (solid line) and the long-wavelength 
CF exciton (dashed line) for heterojunction as a function of density, 
$\rho$.  The typical uncertainty in the theoretical results is shown at the
beginning of each curve.  Experimental energies are also shown, taken from
Ref.~\protect\cite{Davies} (circle), Ref.~\protect\cite{Mellor} 
(diamond), Ref.~\protect\cite{Zeitler}
(square), and Ref.~\protect\cite{Devitt} (down-triangle); 
the filled symbols correspond to the roton, and the empty ones to the
long wavelength mode.
\label{fig2}}
\end{figure}

\begin{figure}
\centerline{\psfig{figure=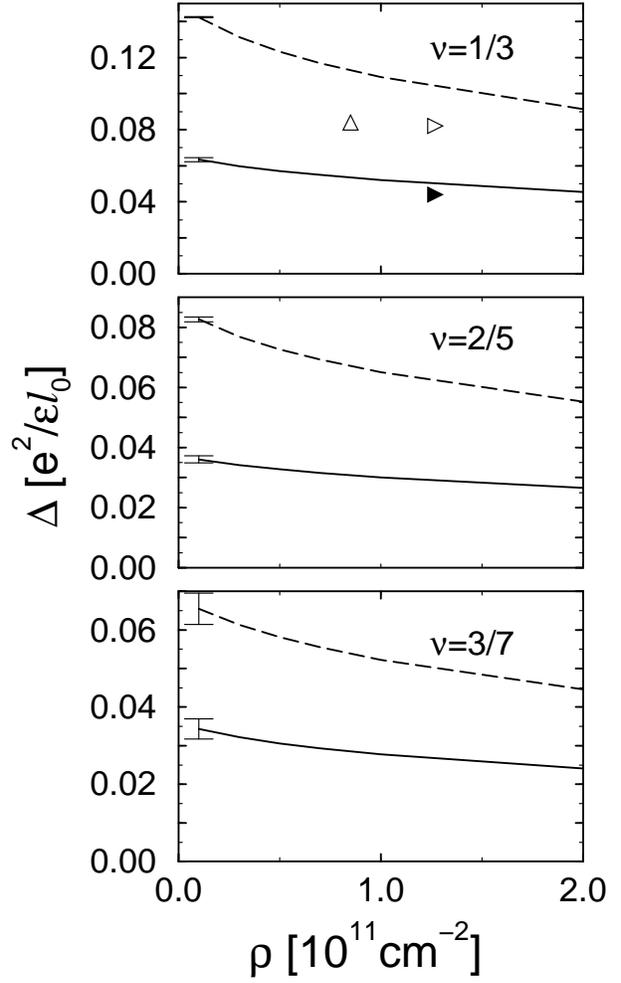,width=5.0in,angle=0}}
\caption{Same as in Fig.~(\protect\ref{fig2}) but for 
square quantum well of width 25nm.  The experimental results are taken from 
Ref.~\protect\cite{Pinczuk1} (up-triangle) 
and Ref.~\protect\cite{Pinczuk2} (right-triangle);
the filled (empty) symbols correspond to the roton (long wavelength mode).
\label{fig3}}
\end{figure}

Figs.~(\ref{fig2}) and (\ref{fig3}) plot the energy of the CF roton and the
long wavelength CF exciton for a heterojunction and 
a square quantum well (of width 25nm) 
as a function of density, calculated with the realistic LDA interaction,
along with the experimental energies obtained in phonon absorption
(at 1/3, 2/5, and 3/7)
as well as in inelastic light scattering experiments (at 1/3). 
A more detailed comparison is given in Table II.
In the small wave vector limit, the calculated  
energy at $1/3$ is off by $\sim$ 30\%.  It has been  
suggested that here the true lowest energy excitation may 
contain {\em two} CF-excitons, and there has been debate as to which 
excitation is being probed by the Raman
scattering in this case \cite{GMP,He}.  For the roton, the 
theoretical energies, obtained with no adjustable parameters, 
are in excellent agreement with the observed ones \cite {Comment}.
One may worry that the situation will be spoiled by Landau level mixing.
This turns out not to be the case.
Following Ref.~\cite {Bonesteel1}, we have estimated the importance 
of LL mixing for the roton energy by considering a 
variational wave function which is a linear combination of the projected 
and unprojected wave functions, and found that the corrections are 
on the order of 5\% for typical densities, consistent with similar  
conclusion for the transport gap \cite {Bonesteel1,Bonesteel2}.

While the ballistic phonon absorption experiments directly measure the 
minimum energy (i.e., the roton) \cite{Mellor2}, the Raman experiments 
ideally probe the $kl_0\rightarrow 0$ limit of the CF
exciton dispersion, the wavelength of the light being much larger 
than $l_0$.   However, a breakdown of momentum conservation due to 
the presence of disorder can  
activate rotons as well \cite{Pinczuk4}, as a result of a singularity in the
density of states.  This has been crucial in explaining multiple peaks in the
Raman spectra for the inter-LL excitations.   
At $\nu=1/3$, a low energy Raman peak has been interpreted as the roton \cite
{Pinczuk2,Davies}.
Recently, Kang {\em et al.} \cite {Pinczuk3} have also observed modes at 
2/5 and 3/7, at energies of 0.031 and 0.008
$e^2/\epsilon l_0$, respectively, which they interpret as the long wavelength 
neutral mode.  Our calculated energies at 2/5 and 3/7 for a quantum
well of width 30nm and density $\rho=5.4\times 10^{10}$ cm$^{-2}$ are 
0.031(3)  and 0.021(3)  $e^2/\epsilon l_0$, respectively, for the roton,  
and 0.070(1) and  0.056(3) $e^2/\epsilon l_0$ for the $kl_0=0$ limit of 
CF exciton.  At 2/5, the energy of the observed excitation is consistent 
not only with the calculated roton energy but also with those measured in 
ballistic phonon absorption experiments, and substantially smaller 
than the calculated $kl_0\rightarrow 0$ limit, which 
might suggest an identification with the roton.  [We note here that 
the observation of the 1/3 roton implies that the violation of the
momentum conservation is sufficiently wide-spread as to render the fundamental
rotons at 2/5 and 3/7 observable as well, which occur at roughly the
same wave vectors ($k l_0\approx 1.6-1.7$) as the 1/3 roton
($k l_0\approx 1.4$).] The energy of the 3/7 mode of Ref.~\cite{Pinczuk3} is 
anomalously low, however.  Further work will be required to 
ascertain the origin of these new Raman 
modes; an experimental observation of 
multiple roton peaks will be especially helpful in clarifying this issue.

In conclusion, the insensitivity of the roton energy to disorder has 
afforded an opportunity for a direct quantitative confirmation of our 
theoretical understanding of the excitations of the fractional quantum Hall effect.
We are grateful to Professor Aron Pinczuk for communicating his 
results to us prior to publication and for the continuous exchange of 
information, and to Xiaomin Zu for numerous helpful discussions.  This work 
was supported in part by the National Science Foundation under grant no. 
DMR-9615005, and by a grant of computing time by the National Center for 
Supercomputing Applications at the University of Illinois (Origin 2000).

\pagebreak

\begin{table}[t]
\caption{Energies of the CF roton and the long wavelength neutral excitation 
at 1/3, 2/5, and 3/7, for a strictly two-dimensional system, in units 
of $e^2/\epsilon l_0$.  
Also given is an estimate for the CF roton ``mass", $m_R^*\equiv \mu_R^* m_e
\sqrt{B[T]}$, defined in Eq.~(\ref{mass}), where $m_e$ is the electron mass in
vacuum.  The statistical uncertainty in the last
digit(s) is shown in parentheses.
\label{tab:Tab1}}
\vspace{0.4cm}
\begin{center}
\begin{tabular}{|c|c|c|c|} 
mode & $\nu$ & energy &  $\mu_R^*$  \\ \hline \hline
$kl_0=0$ & 1/3  & 0.15 &  -  \\ \cline{2-4}
& 2/5  & 0.087(1) &   - \\ \cline{2-4}
& 3/7  & 0.068(5) &  - \\ \hline \hline
roton &1/3  & 0.066(1) &  0.0079(3) \\ \cline{2-4}
& 2/5  & 0.037(1) &  0.0090(11) \\ \cline{2-4}
& 3/7  & 0.027(3) &  0.0095(32) \\ 
\end{tabular}
\end{center}
\end{table}

\begin{table}[t]
\caption{Comparison of theory and experiment for the roton energy as well as
the energy of the long-wavelength exciton, quoted in units of $e^2/\epsilon
l_0$.
In Ref.~\protect\cite{Mellor}, the roton energies were determined for
2/3, 3/5, and 4/7, which, assuming particle-hole symmetry, are the same as
the roton energies at 1/3, 2/5, and 3/7, when measured in units of
$e^2/\epsilon l_0$. 
\label{tab:Tab3}}
\vspace{0.4cm}
\begin{center}
\begin{tabular}{|c||c|c||c|c||c|} 
$\nu$ & \multicolumn{2}{c||}{$kl_0=0$} &  \multicolumn{2}{c||}{roton} & 
 Reference \\ \cline{2-5} 
& experiment& theory   & experiment& theory   &   \\  \hline
1/3 & 0.082 & 0.104(1) & 0.044     & 0.050 (1)& \protect\cite{Pinczuk2} 
\\  \cline{2-6}
    & 0.084 & 0.113(1) & -         & 0.052(1) & \protect\cite{Pinczuk1} 
\\  \cline{2-6}
    & -     & 0.09(2)  & 0.041(2)  & 0.045(1) & \protect\cite{Mellor}
 \\  \cline{2-6}
    & 0.074 & 0.095(1) & 0.047     & 0.047(1) & \protect\cite{Davies}
 \\  \cline{2-6}
    & -     & 0.092(1) & 0.036(5)  & 0.045(1) & \protect\cite{Zeitler}
 \\  \hline \hline
2/5 & -     & 0.054(1) & 0.021(2)  & 0.026(1) & \protect\cite{Mellor}
 \\  \cline{2-6}
    & -     & 0.055(1) & 0.025(3)  & 0.027(1) & \protect\cite{Devitt}
 \\  \hline \hline
3/7 & -     & 0.044(2) & 0.014(2)  & 0.017(2) & \protect\cite{Mellor}
 \\ 
\end{tabular}
\end{center}
\end{table}

\end{document}